\documentclass{article}
\usepackage{amsmath}
\usepackage{graphicx}
\input{tcilatex}
\begin{document}

\title{\textquotedblleft Instantaneous superluminality\textquotedblright\ in
a bimetallic wire consisting of a superconducting aluminum wire plated with
a thick copper covering }
\date{White paper of 11-14-10 by R. Y. Chiao\thanks{%
U. C. Merced, e-mail address: rchiao@ucmerced.edu}}
\author{}
\maketitle

\begin{figure}
\includegraphics[width=6in]{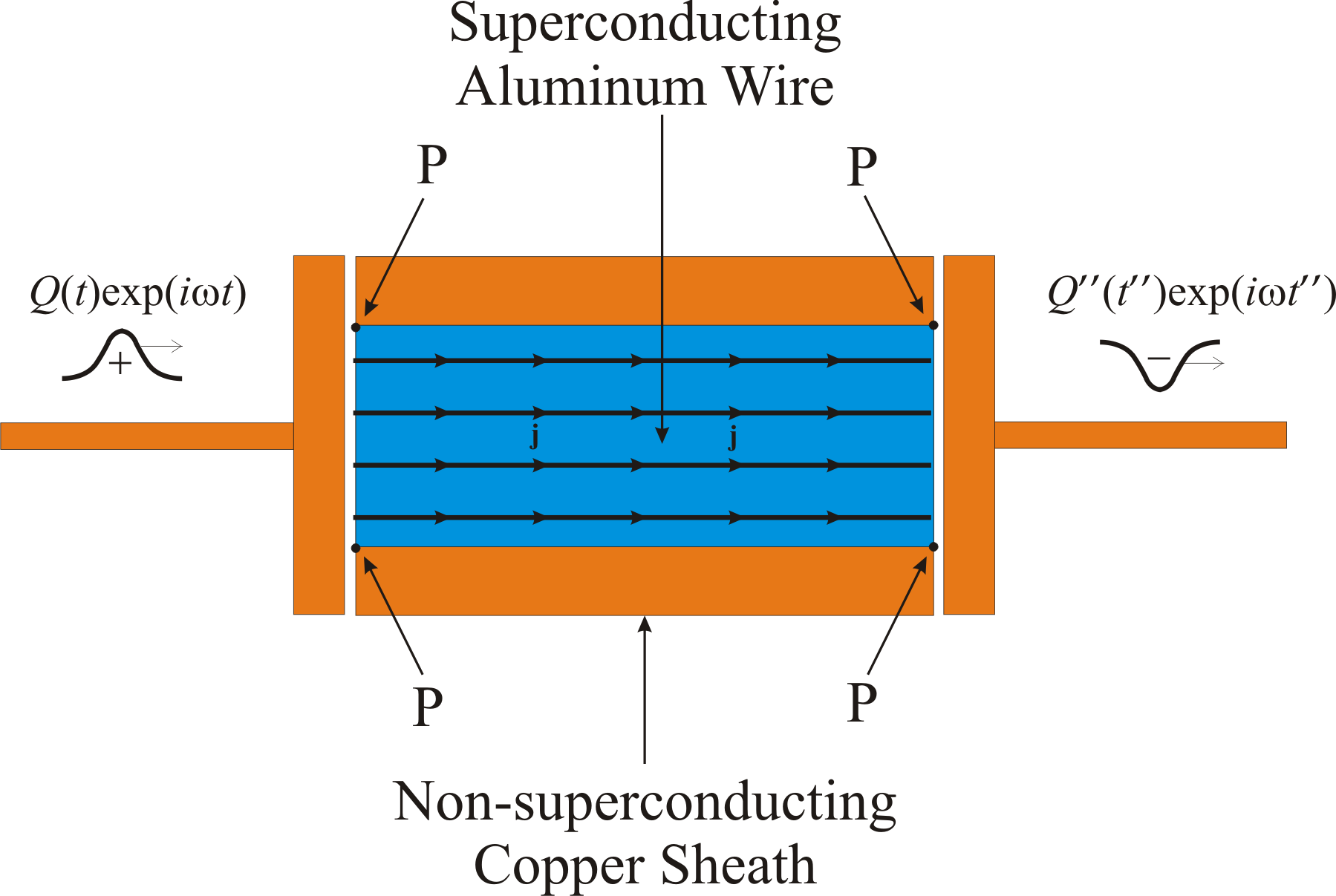}
\caption{(Not to scale). A simple model
of a \textquotedblleft bimetallic wire\textquotedblright , consisting of an
aluminum wire overcoated by a thick copper sheath, for demonstrating the
possibility of \textquotedblleft instantaneous
superluminality\textquotedblright . A SC aluminum wire (in \textit{blue}) is
covered with a thick NSC copper sheath (in \textit{orange}). The
equally-spaced horizontal streamlines inside the SC portion of the wire
denote the supercurrent density $\mathbf{j}$. The wire is excited by an
charge pulse $Q(t)\exp (i\protect\omega t)$ coming in from the left, and the
wire transmits a superluminal outgoing pulse $Q^{\prime \prime }(t^{\prime
\prime })\exp (i\protect\omega t^{\prime \prime })$ to the right. Points P
are places where charges partition into luminal (\textquotedblleft type
(i)\textquotedblright ) surface charges, and superluminal (\textquotedblleft
type (ii)\textquotedblright ) volume charges \protect\cite{Partition}.}
\end{figure}

We shall analyze here the transmission of microwave-frequency electrical
pulses through a \textquotedblleft bimetallic\textquotedblright\ wire
consisting of superconducting aluminum (Al) wire covered with a thick,
nonsuperconducting copper (Cu) sheath that is many skin depths thick. See
Figure 1. This is another simple example that will demonstrate the
possibility of \textquotedblleft instantaneous
superluminality\textquotedblright\ \cite{Prague}\ within a \textquotedblleft
bimetallic\textquotedblright\ superconducting system \cite{e-beam}.

The analysis starts from Maxwell's modification of Ampere's law by the
displacement current%
\begin{equation}
\nabla \times \mathbf{B}=\mu _{0}\left( \mathbf{j}+\frac{\partial \mathbf{D}%
}{\partial t}\right) \text{ ,}  \label{Ampere's law}
\end{equation}%
where $\mu _{0}$ is the magnetic permeability of free space, $\mathbf{B}$ is
the magnetic induction, $\mathbf{j}$ is the supercurrent density of Cooper
pairs at an arbitrary point within the Al wire, and $\mathbf{D}$ is the
displacement field at the same point. We shall examine circumstances under
which the supercurrent density within the Al wire portion of the
bimetallic-wire configuration can be cancelled out by the displacement
current density, such that%
\begin{equation}
\mathbf{j}+\frac{\partial \mathbf{D}}{\partial t}=\mathbf{0}\text{
everywhere inside the Al wire .}  \label{Cancellation of currents}
\end{equation}%
We shall call this the \textquotedblleft cancellation-of-currents
condition.\textquotedblright\ This cancellation-of-currents condition (\ref%
{Cancellation of currents}) is consistent with the continuity equation,
since if one were to take the divergence of (\ref{Cancellation of currents}%
), one obtains%
\begin{equation}
\nabla \cdot \mathbf{j}+\frac{\partial \left( \nabla \cdot \mathbf{D}\right) 
}{\partial t}=\nabla \cdot \mathbf{j}+\frac{\partial \rho }{\partial t}=0%
\text{ .}  \label{continuity equation}
\end{equation}%
It is clear by inspection by the Maxwell equation (\ref{Ampere's law}) that
when condition (\ref{Cancellation of currents}) holds, the sources of the
magnetic induction $\mathbf{B}$, which is generated by the point-by-point 
\emph{superposition} of the supercurrent density $\mathbf{j}$\ and the
displacement current density $\partial \mathbf{D/}\partial t$, will vanish,
since these two sources for generating the $\mathbf{B}$ field will cancel
each other, and therefore that%
\begin{equation}
\nabla \times \mathbf{B}=\mathbf{0}\text{ ;}  \label{curl B = 0}
\end{equation}%
\begin{equation}
\nabla \cdot \mathbf{B}=0\text{ .}  \label{div B = 0}
\end{equation}%
Note that these two equations will be valid even when the supercurrent
density $\mathbf{j}$ and the displacement current density $\partial \mathbf{D%
}/\partial t$\ may be oscillating at microwave frequencies.

We shall show below that the cancellation-of-currents condition (\ref%
{Cancellation of currents})\ can occur when a superconducting Al wire is
coated by a thick nonsuperconducting Cu sheath, as indicated in Figure 1, so
that there is an intimate electrical contact between the Al and Cu metals at
their interface. This can be implemented by copper-plating aluminum wire. We
shall call this a \textquotedblleft bimetallic-wire
configuration.\textquotedblright\ The Cu coating on the surface of the Al
wire will be chosen to be many skin depths thick, so that at microwave
frequencies, currents can flow only on the outer surface of the Cu sheath
overlaying the Al wire, and thus these normal electrical currents will be
far removed from the interface between the Cu and Al, i.e., they will remain
far away from the surface of the Al wire. As a result, any high-frequency $%
\mathbf{B}$ field will decay exponentially away from the outer surface of
the Cu sheath on the scale of the microwave skin depth of Cu, which is on
the scale of microns, and therefore will become vanishingly small at the
interface between the Al and the Cu. Hence the Cu sheath effectively forms a
tightly fitting Faraday cage over the Al wire.

Therefore the boundary conditions at the surface of the Al wire will be such
that a $\mathbf{B}$ field oscillating at microwave frequencies effectively
vanishes at the interface between the Al and the Cu metals. Otherwise, by
Faraday's law of induction, any such high-frequency $\mathbf{B}$ fields at
the interface would generate high-frequency $\mathbf{E}$ fields within the
Cu sheath. However, such electric fields will be shorted out by the high
electrical conductivity of the Cu metal immediately surrounding the Al wire.
Hence the boundary conditions for the $\mathbf{B}$ field on the interface
between the Al and the Cu, i.e, on the surface of the Al wire, are, to a
very good approximation,%
\begin{equation}
\mathbf{B}=\mathbf{0}\text{ everywhere on the surface of the Al wire ,}
\end{equation}%
and therefore from Maxwell's equations (\ref{curl B = 0}) and (\ref{div B =
0}), it follows that%
\begin{equation}
\mathbf{B}=\mathbf{0}\text{ everywhere inside the volume of the Al wire .}
\label{B=0 in volume of wire}
\end{equation}%
However, from the fact that the supercurrent $\mathbf{j}$ does not vanish
inside the Al wire, it follows that charges must accumulate at the ends of
the wire, so that a displacement $\mathbf{D}$\ field must be set up \textit{%
within} the wire. This displacement field satisfies the Maxwell equation%
\begin{equation}
\nabla \cdot \mathbf{D}=\rho   \label{div D = rho}
\end{equation}%
where $\rho $ is the non-vanishing charge density of Cooper pairs
accumulating at the two opposite ends of the wire. Note that $\rho $ and $%
\mathbf{D}$ will be oscillating at microwave frequencies.

In order to show that the cancellation-of-currents condition (\ref%
{Cancellation of currents}) is satisfied inside the bimetallic-wire
configuration, let us analyze the equations obeyed by the supercurrent
density $\mathbf{j}$ within the Al wire portion of this configuration. Let
us define a time-dependent superflow velocity field $\mathbf{v}$ of the
Cooper pairs \ within the superconductor through the relationship%
\begin{equation}
\mathbf{j}=\rho \mathbf{v}\text{ ,}
\end{equation}%
where the charge density $\rho $ is related to the complex order parameter $%
\psi $ as follows:%
\begin{equation}
\rho =q\psi ^{\ast }\psi
\end{equation}%
where $q$ is the charge of a Cooper pair.

In order to avoid the enormous Coulomb energies associated with any possible
unbalanced charge densities arising from inhomogeneities in the Cooper pair
density inside the Al wire, one demands that the charge density of the ionic
lattice of the superconductor must be \textit{exactly} compensated by the
charge density of the Cooper pairs at every point inside the volume of the
superconductor away from the surface. Since we will also assume that the
ionic lattice possesses a constant, \textit{homogeneous} density everywhere
inside the Al wire, it follows that the Cooper pair charge density $\rho $
must be a constant of the motion, i.e.,%
\begin{equation}
\rho =q\psi ^{\ast }\psi =\text{ constant .}
\end{equation}%
This is consistent with the fact that the ground BCS state of the
superconductor corresponds to a uniform charge-density state, and the fact
that in first-order perturbation theory, the ground state remains unaltered
to lowest order by external perturbations. It follows that%
\begin{equation}
\frac{\partial \rho }{\partial t}=0\text{ within the volume of the Al wire .}
\end{equation}%
Therefore from the continuity equation (\ref{continuity equation}), it
follows that%
\begin{equation}
\rho \nabla \cdot \mathbf{v}+\frac{\partial \rho }{\partial t}=\rho \nabla
\cdot \mathbf{v}=0
\end{equation}%
and therefore that%
\begin{equation}
\nabla \cdot \mathbf{v}=0\,\ .
\end{equation}%
One concludes that the superflow velocity field $\mathbf{v}$\ of the Cooper
pairs inside the Al wire of the bimetallic-wire configuration is \textit{%
incompressible}.

From DeWitt's minimal coupling rule \cite{Prague}, we showed that%
\begin{equation}
\mathbf{v}=-\frac{q}{m}\mathbf{A}-\mathbf{h}\text{ ,}
\label{superfluid velocity in terms of A and h}
\end{equation}%
where $\mathbf{v}$ is the superfluid velocity field, $q$ is the charge and $m
$ is the mass of the Cooper pair, respectively, $\mathbf{A}$ is the vector
potential, and $\mathbf{h}$ is DeWitt's vector potential. The DeWitt (or
\textquotedblleft radiation\textquotedblright ) gauge is being assumed here,
with%
\begin{equation}
\nabla \cdot \mathbf{h}=\nabla \cdot \mathbf{A}=0\text{ .}
\end{equation}%
Taking the curl of the superfluid velocity field given by (\ref{superfluid
velocity in terms of A and h}), one obtains, to a very good approximation,%
\begin{equation}
\nabla \times \mathbf{v}=-\frac{q}{m}\nabla \times \mathbf{A}-\nabla \times 
\mathbf{h}=-\frac{q}{m}\mathbf{B}-\mathbf{B}_{\text{G}}\text{ }=\mathbf{0}%
\text{,}
\end{equation}%
where $\mathbf{B}=\nabla \times \mathbf{A}=\mathbf{0}$ is the magnetic
field, which vanishes by the solution (\ref{B=0 in volume of wire}), and $%
\mathbf{B}_{\text{G}}=\nabla \times \mathbf{h}=\mathbf{0}$ is the
gravito-magnetic field \cite{Prague}, which we shall also assume to vanish
everywhere inside the Al wire.

Therefore the superflow velocity field $\mathbf{v}$ for Cooper pairs inside
the Al wire obeys the two equations%
\begin{eqnarray}
\nabla \times \mathbf{v} &=&\mathbf{0}\text{~;}  \label{curl v = 0} \\
\nabla \cdot \mathbf{v} &=&0\text{ .}  \label{div v = 0}
\end{eqnarray}%
One concludes that the superflow of Cooper pairs inside the Al wire of the
bimetallic-wire configuration shown in Figure 1 is both \textit{irrotational}
and \textit{incompressible}. It follows from (\ref{curl v = 0}) that a
solution exists of the form%
\begin{equation}
\mathbf{v=\nabla }\varphi 
\end{equation}%
for some potential function $\varphi $, and therefore from (\ref{div v = 0})
that%
\begin{equation}
\mathbf{\nabla }^{2}\varphi =0\text{ .}
\end{equation}%
Thus $\varphi $ obeys Laplace's equation, i.e., the Cooper-pair superflow is 
\textit{streamline} flow. We know that in the special case of the laminar
superflow within a straight pipe of constant cross section, the streamlines
are the horizontal straight lines indicated in Figure 1, which satisfy the
1D solution of Laplace's equation, i.e.,%
\begin{equation}
\varphi (x,y,z,t)=C(t)x\text{ where }C(t)\text{ is constant independent of }%
x,y,z\text{ ,}  \label{1D solution of Laplace equation}
\end{equation}%
where the $x$ direction has been chosen to coincide with the direction of
the horizontal superflow indicated in Figure 1. The superflow may be \textit{%
time-dependent} due to the time variations of the incident charge pulse $%
Q(t)\exp (i\omega t)$ coming in from the left; nevertheless, there will
exist \textit{instantaneous} streamline solutions \textit{everywhere} inside
the Al metal having the form given by (\ref{1D solution of Laplace equation}%
).

Therefore the solution for the supercurrent $\mathbf{j}$ inside the Al wire
of the bimetallic-wire configuration shown in Figure 1\ is given by%
\begin{equation}
\mathbf{j}(t)=\mathbf{\hat{\imath}}\rho C(t)\text{ everywhere inside the Al
wire ,}  \label{solution for j}
\end{equation}%
where $\mathbf{\hat{\imath}}$ is the unit vector in the $x$ direction of the
superflow. As a result, charge will be accumulating at the right end of the
Al wire at a rate%
\begin{equation}
\frac{dQ}{dt}=jA=\rho C(t)A\text{ ,}
\end{equation}%
where $j$ is the $x$ component of the supercurrent, and $A$ is the
cross-sectional area of the Al wire. By charge conservation, there must
exist an equal but oppositely signed charge accumulating at the left end of
the Al wire. This will produce a uniform displacement field inside the Al
wire.

From the divergence theorem and Gauss's law, i.e., starting from Maxwell's
equation%
\begin{equation}
\nabla \cdot \mathbf{D}=\rho 
\end{equation}%
and applying a \textquotedblleft pillbox\textquotedblright\ argument to the
right-end surface of the Al wire \cite{Normal component of D is continuous},
it follows that there will be an $x$ component displacement field $D$, with
the vector field $\mathbf{D}$ directed along the $-x$ axis, produced by the
charge $+Q$ accumulating on the right end of the wire, and, by a similar
argument, of the charge $-Q$ accumulating on the left end of the wire, which
is given by%
\begin{equation}
D=-\sigma =-\frac{Q}{A}=-\rho \dint C(t)dt\text{ .}
\end{equation}%
at each instant of time $t$. Taking the time derivative of this displacement
field and comparing it with (\ref{solution for j}), one obtains the
following relationship for the $x$ components of the supercurrent and the
displacement current density: 
\begin{equation}
j=-\frac{\partial D}{\partial t}\text{ .}
\end{equation}%
Therefore, we have verified that, in this special case, the
cancellation-of-currents condition (\ref{Cancellation of currents}) is
indeed satisfied by the bimetallic-wire configuration shown in Figure 1.

In summary, the only places in the Al wire where charge can accumulate, and
therefore where the charge density can change with time, is either at the
left end of the wire, where the charge from the incoming Gaussian-envelope
microwave pulse is being induced, or at points at the right end of the wire,
where this induced charge re-appears in just such way that the \textit{total}
charge of the entire Al wire portion of the bimetallic-wire configuration is
always \textit{exactly} conserved at every single instant of time $t$. This
implies that the disappearance of a given electron at the left end of the Al
wire$\,$is always accompanied by its \textit{simultaneous} reappearance at
an arbitrarily far-away point at the right end of the Al wire at \textit{%
exactly} the same instant of time $t$. Otherwise, the principle of charge
conservation would be violated.

Following \cite{e-beam}, we shall call this counter-intuitive effect
\textquotedblleft instantaneous superluminality within a bimetallic
superconducting wire.\textquotedblright\ It can be easily verified or
falsified in a modification of the simple experiment described in \cite{coax}%
, except here one can eliminate the dielectric and the outer conductor of
the superconducting-core coaxial cable, and use instead a long coil of bare,
copper-plated aluminum wire. Also, one can simply implement the figure `8'
gravitational-wave antenna configuration presented in \cite{e-beam} by
stripping away the outer layers of the coaxial cable bent into the shape of
a figure `8', thus leaving only the bimetallic-wire central core of this
cable as the antenna.

Again, it should be emphasized that this superluminal effect does not
violate relativistic causality because the incident charge pulse has an 
\textit{analytic} waveform, for example, a Gaussian pulse, with a finite
bandwidth (i.e., with frequencies less than the BCS gap). There exists no
discontinuous \textquotedblleft front\textquotedblright\ within the Gaussian
waveform, before which the waveform is \textit{exactly} zero. Such a
\textquotedblleft front\textquotedblright\ would contain infinitely high
frequency components that would exceed the BCS gap frequency, and thus
destroy the superconductivity of the wire. Again, this \textquotedblleft
instantaneously superluminal\textquotedblright\ effect has similarities with
that of a Gaussian wavepacket tunneling through a tunnel barrier in quantum
mechanics, whose early analytic tail contains all the information needed to
reconstruct the entire transmitted wave packet, including its peak, earlier
in time \textit{before} the incident peak could have arrived at a detector
traveling at the speed of light \cite{Chiao-Steinberg}.

\end{document}